\documentclass{article}
\begin{document}

\title{Derivation of the Quantum Probability Rule without the Frequency	Operator}
\author{Sumio Wada \\ Institute	of Physics \\ The University of	Tokyo, Komaba \\ Tokyo, 153-8902, Japan \\ \thanks{Electric	address: wada@hep1.c.u-tokyo.ac.jp}}
\maketitle

\begin{abstract}
We present an alternative frequencists'	proof of the quantum probability rule which does not make use of the frequency operator, with expectation that this	can circumvent the recent criticism against the	previous proofs	which use it.  We also argue that avoiding the frequency operator is not only for technical	merits for doing so but	is closely related to what quantum mechanics is	all about from the viewpoint of many-world interpretation.

\begin{flushleft}
\end{flushleft}

\end{abstract}

\section{Introduction}

Whether	the quantum probability	rule (QPR), or Born's rule, should be postulated as one of axioms of quantum mechanics has been a long standing issue since	the authors of refs.1) and 2) claimed that this	rule can be derived by reinterpretting the probability as the frequency of outcomes	of measurements. They introduced a so-called frequency operator, which expresses the relative frequency (the proportion of the frequency) of a certain specific outcome in the infinite repetition of measurements,	and claimed that a infinite repetition state is an eigenstate of this operator with	the eigenvalue which QPR dictates.  However it was argued that their statement is about	many but not infinite repetition of measurements and did not constitute a rigorous proof that the infinite repetition state	is an eigenstate. Their	proof was claimed to correspond	to the weak law of large number in the classical probability theory.

In refs.3) and 4), the authors improved	this situation.	They defined the frequency operator	acting on the infinite repetition state, and proved what amounts to the	strong law of large number.  But in the	course of the proof they had to introduce the measure, i.e.	the weight of each infinite sequence of	outcomes. Recently, in ref.5), it was pointed out that the measure and so the frequency operator itself is not unique and, as a result, we can arbitrarily changes its eigenvalue.	Adopting a specific form of the	measure	amounts	to adopting an assumption about the	probability rule from the outset. If so, the claim that	QPR can	be derived from	the frequency becomes dubious.

What we	try to do in this paper	is to prove QPR, sticking to frequencists' approach but without recourse to	the frequency operator.	 We circumvent the above criticism by avoiding the frequency operator, but we also take	note that there is a more fundamental reason to avoid the frequency	operator.

In the orthodox	formulation of quantum mechanics, beside the probability rule, the following ansatz	is often postulated.

\begin{quotation}
Postulate: Every observable is associated with a Hermitian operator $O$. Only possible outcomes of a measurement of	$O$ are	eigenvalues of $O$. When a system is in	an eigenstate $|\psi>$ of $O$, i.e. $O|\psi> = \lambda |\psi>$,	then a measurement of $O$ will yield the value $\lambda$.
\end{quotation}

This is	the reason why the frequency operator F	was introduced.	If F were unambiguously defined and	a certain state	were an	eigenstate of the F with the eigenvalue	p, then	it would follows from the above	that the relative frequency in which a certain outcome occurs would	be p.

I challenge this type of reasoning from	the viewpoint of a many-worlder	and quantum cosmologist. Any measurement is	a phenomenon which occurs in the Universe, and as such,	should be expressed in the state of the	world (i.e. in the wave function of	the Universe).	Mathematically we may be able to devise	an operator by which we	can investigate	a property of the state, but that should not be required.

As a simple example, suppose that a wave function $\psi(q)$ has	a form
\begin{equation}
	    \psi(q)  \propto \delta(q-q_0)
\end{equation}
where q	is a c-number variable and $q_0$ is a certain constant.	This wave function vanishes	unless $q=q_0$ and so it should	be natural to claim that $q=q_0$ in this state.	

However, if we stick to	the above postulate, we	have to	introduce the operator $\hat{q}$ which satisfies
\begin{equation}
  \hat{q}\,\psi(q) =   q_0\,\psi(q)
\end{equation}
and deduce that	the outcome of measurements of the observable $\hat{q}$	should be $q_0$. One may wonder whether this procedure is really necessary.	 If we are concerned with the outcome of the measurement, then we introduce the	wave function $\Psi$ of the	apparatus whose	readout	$Q$ is in one-to-one correspondence with $q$. Then, if the measurement is accurate, we will get
\begin{equation}
   \Psi(Q)  \propto  \delta(Q-Q_0)
\end{equation}
from which we should be	able to	deduce that $Q=Q_0$ (and so $q=q_0$) without recourse to any operator. 

In the above simple example the	distinction between eq.(1) (or eq.(3)) and eq.(2) looks to be practically irrelevant.	 However, when we seriously ponder over	what wave functions actually mean and, in particular, when the definition of the operator itself is	ambiguous, the distinction may not be an irrelevant issue anymore.

In this	paper we insist	that the place to look for predictions in quantum theory is not an operator	(nor an	observable) but	a wave function.  An operator is regarded as no	more than a means to extract information when convenient. It is from this viewpoint	that we	reexamine frequencists'	derivation of QPR and try to derive the	same conclusion	without	recourse to the	frequency operator.

\section{Naive Derivation}

The difficulty in this problem lies in deriving	information about the infinite repetition of measurements from considerations on many but finite repetition	of measurements.  In this section we present a "proof" of QPR without a	frequency operator but will	take a rather naive attitude on	this point, optimistically assuming that a limit of finite repetitions might give the required result about the infinite repetition.  (The calculation shown below was presented before in ref.6).

We start from a	finite repetition state
\begin{equation}
     |\Psi>_N  =  |\psi,1>|\psi,2>|\psi,3>\dots|\psi,N>
\end{equation}
Each element $|\psi,i>$	is supposed to represent a single identical particle with spin, each of which is prepared in a spatially different position	but is supposed	to be in an identical state as long as the spin	is concerned.  When this particle is a fermion, the	whole state should be antisymmetrized, but this	requirement is assumed to be taken care	of by the spatial part of the wave function.  The spin part	of the wave function, which is of our concern in the following argument, is symmetric because all the spin states are identical.  For simplicity, $|\Psi>_N$ is supposed to	represent only the spin	part of	the state.

Because	all the	spins are assumed to be	in an identical	state, we can write as
\begin{equation}
    |\psi,i>  =	 a |\alpha,i> +	b |\beta,i>
\end{equation}
where ƒ¿ and ƒÀ	represent a spin state while $a$ and $b$ are $i$-independent coefficients with
\begin{equation}
    |a|^2  + |b|^2  =  1 
\end{equation}

We substitute eq.(5) into eq.(4) and expand it.	 We get	various	terms, each of which has certain numbers of	$|\alpha>$ and $|\beta>$ . When	the number of $|\alpha>$ is $n$, the number of $|\beta>$ is $N-n$ and the relative frequency r (of $|\alpha>$) is defined as
\begin{equation}
	      r	 =  n/N
\end{equation}

Next we	introduce the normalized and symmetrized state $|r>$ with a fixed relative frequency, which	is
\begin{equation}
      |r>_N \equiv _NC_n^{-1/2}	\{|\alpha,1> \dots |\alpha,n>|\beta,n+1> \dots |\beta,N> + perm.\}
\end{equation}
The set	$\{|r> : r=n/N,	n=1\sim	N\}$ is	complete in the	space of symmmetrized $N$-copy states, and we can expand $|\Psi>_N$	in terms of $|r>$ as
\begin{equation}
	|\Psi>_N  =  \sum_r c_N(r) |r>_N 
\end{equation}
where
\begin{equation}
	c_N(r)	=   _NC_n^{1/2}	a^n b^{N-n}
\end{equation}
It is easy to show that
\begin{equation}
	  \sum |c_N(r)|^2  =  1
\end{equation}
\begin{equation}
     \sum (r-|a|^2)^2\,|c_N(r)|^2 = |a|^2|b|^2/N ( \to 0\, as\,	N \to \infty)
\end{equation}
These results imply that $|c_N(r)|^2$ concentrates at $r = |a|^2$ in the limit $N \to \infty$.

Note in	passing	that the above calculation with	the two-level systems can be easily generalized to multi-level systems and also to continuous variable cases, as are shown	in Appendices.

\section{An Alternative	Postulate of Quantum Mechanics}

The above results tempt	us to conclude that the	relative frequency of the outcome ƒ¿ in the	infinite repetition of spin measurements is $|a|^2$, as	QPR dictates.  If	this claim can be justified, we	need not to include QPR	in the set of axioms (postulates)	of quantum mechanics, although we do need some alternative postulate, which will be weaker than QPR, to justify the above	argument.  In this section we investigate	what we	actually need, propose a weaker	postulate and modify the above argument accordingly.

We start from a	trivial	case in	which the wave function	$\psi$ is a function of a discrete variable	$n$ and	is proportional	to Kronecker's delta,
\begin{equation}
    \psi(n)  \propto  \delta_{nn_0}
\end{equation}
Naturally we conclude that the value of	$n$ in this state is $n_0$. Its	extension to a continuous variable (say $q$) might be
\begin{equation}
    \psi(q)   \propto  \delta(q-q_0)
\end{equation}
in which case we would claim that $q=q_0$. However this	reasoning can not be applied to $c(r)$ (which is the $N\to\infty$ limit of $c_N(r)$	in the previous	section).  While $\delta(q-q_0)$ is not	normalizable  (not square-integrable),  $c(r)$ was normalized.  What behaves like a	delta function is not $c(r)$ but $|c(r)|^2$. 

Normalizable functions are more	legitimate states in an	ordinary rigorous treatment of quantum mechanics, and so, any postulate of the quantum theory should be at least applicable	to them.  It should also be applicable whether a variable is discreet or continuous. We will pursue	such an	extension of eq.(13).

First we note that, judging from the situation for $c(r)$, natural extension of eq.(13) to continuous variable cases should	not be eq.(14),	but something which involves $|\psi(q)|^2$.  (Note that	 $\psi(n) \propto \delta_{nn_0}$ and $|\psi(n)|2  \propto \delta_{nn_0}$ are equivalent in discreet	variable cases.)

Let us start from a simple example.  We	consider a space of functions ${\psi(q)}$ in which the inner product is written as
\begin{equation}
   \sum	 \psi_1(q)^\ast	\cdot \psi_2(q)
\end{equation}
A variable $q$ is assumed to be	real and the summation is over all possible values of $q$. $q$ can be either discrete or continuous.  When $q$ is continuous, the summation	in the above equation is replaced with an integral.

For later convenience we introduce the following notation
\begin{equation}
	\rho_q(\psi_1,\psi_2)	\equiv	\psi_1(q)^\ast \cdot \psi_2(q)
\end{equation}
\begin{equation}
	\rho_q(\psi)  \equiv  |\psi(q)|^2
\end{equation}
$\rho_q(\psi)$ is nothing but $\rho_q(\psi_1,\psi_2)$ with $\psi_1 = \psi_2 =\psi$.

$\psi(q)$ can be interpreted as	a $q$-representation of	a state	$|\psi>$ , that is
\begin{equation}
     \psi(q) =	<q|\psi>
\end{equation}
and
\begin{equation}
      \rho_q(\psi_1,\psi_2) = <\psi_1|q><q|\psi_2>
\end{equation}
Then the inner product eq.(15) means 
\begin{equation}
       <\psi_1|\psi_2>	\equiv	\sum  <\psi_1|q><q|\psi_2>
\end{equation}
which implies the completeness relation
\begin{equation}
      1	=  \sum	|q><q|
\end{equation}

We assume that a quantum state $|\psi>$	is represented by a normalized vector,. which should be one	of the axioms of quantum theory. This means
\begin{equation}
    \sum \rho_q(\psi) =	<\psi|\psi> = 1
\end{equation}
Suppose	that for any positive number ƒÃ
\begin{equation}
    \sum_{q<q_0-\epsilon} \rho_q(\psi) =  0
\end{equation}
\begin{equation}
    \sum_{q<q_0+\epsilon} \rho_q(\psi) =  1
\end{equation}
Then we	naturally conclude that	the value of $q$ in this state is $q_0$. These conditions are essentially equivalent to say	that $\rho_q =\delta_{qq_0}$ or	$\rho_q	=\delta(q-q_0)$, but we	prefer the above expression to avoid arguments about what the delta	function actually means.

In general the above conditions	do not hold and	the state $\psi$ does not take a specific value of $q$.  However we	may try	another	variable, say $\tilde{q}$, in which case we should examine $\rho_{\tilde{q}}(\psi)$, i.e. the $\tilde{q}$ version of $\rho_q(\psi)$. If there is a completeness relation analogous to eq.(21) for $\tilde{q}$
\begin{equation}
     1 = \sum |\tilde{q}><\tilde{q}|
\end{equation}
then we	can follow the same argument above.  We	define $\rho_{\tilde{q}}(\psi)$ and examine	whether	the conditions similar to eqs.(23,24) hold or not for a	certain	value of $\tilde{q}_0$.

However	we can not use this approach either to derive QPR from the result in the previous section.	Suppose	that we	are working in the (non-separable) space spanned by the	infinite repetition states.  The state defined in eq.(8), which has a fixed	relative frequency $r$,	is a state with	a finite number	of copies and so does not belong to this space,	which means that we can	not utilize the completeness relation to define what is supposed to	become $\rho_r(\Psi)$.

However	we can define the quantity which is very close to that.	 Write a state with infinite number	of copies of a two-level system	as
\begin{equation}
      |\Psi>_\infty  =	|\psi,1>|\psi,2>\dots |\psi,N>\times |\psi,N+1>\dots  (\equiv |\Psi>_N\times |\Psi>_{\infty-N})
\end{equation}
$\psi$ in each element does not	need to	coincide with each other at this stage but eventually we restrict $|\Psi>_\infty$ to the inifinite repetition state.

We define the inner product of two of such states as
\begin{equation}
    _\infty<\Psi_1|\Psi_2>_\infty \equiv \prod_i<\psi_1,i|\psi_2,i>
\end{equation}
$|\Psi>_\infty$	is normalized when all of $|\psi,i>'s$ are normalized. When $|\Psi_2>_\infty$ is a linear combination of two such states as
\begin{equation}
    |\Psi_2>_\infty  = a|\Psi_{2a}>_\infty + b|\Psi_{2b}>_\infty 
\end{equation}
then
\begin{equation}
    _\infty<\Psi_1|\Psi_2>_\infty \equiv  a\prod_i<\psi_1,i|\psi_{2a},i> + b\prod_i<\psi_1,i|\psi_{2b},i>
\end{equation}

Next we	define a $r$-dependent bilinear	functional of $|\Psi>_\infty$ as
\begin{equation}
    \rho_{r,N}(\Psi_1,\Psi_2) =	_N<\Psi_1|r>_N\cdot_N<r|\Psi_2>_N\times	_{\infty-N}<\Psi_1|\Psi_2>_{\infty-N}
\end{equation}
where $|r>_N$ is what was defined in eq.(8) in the previous section.

Note that, in the above	equation, $r$ should be	a rational number ($0<r<1$) for which $rN(=n)$ is an integer.  For a general real number $r$ ($0<r<1$)¤ we consider	a sequence $\{r_N: N=1\sim\infty,\,r_NN=integer\}$ in which $r_N$ is the possible rational number which is nearest to $r$ and we define
\begin{equation}
   \rho_r(\Psi_1,\Psi_2) \equiv	lim_{N\to\infty}\,N\rho_{r_N,N}(\Psi_1,\Psi_2)
\end{equation}
We multiplied ƒÏ by $N$	on the right-hand side to take into account the	fact that $r$ is a continuous	variable while $r_N(=n/N)$ is discrete and $\Delta n=N\Delta r$. This	definition makes it possible to	require
\begin{equation}
   \int_{0<r<1}	\rho_r(\Psi_1,\Psi_2)\,dr = lim_{N\to\infty}\sum_{r_N}\rho_{r_N,N}(\Psi_1,\Psi_2)
\end{equation}
For general states $\Psi$, the limit in	eq.(31)	does not necessarily exist, but does exist for the states which we are actually interested in, namely the infinite repetition states.  In the following we restrict	$\Psi$ to these	states (and their linear combinations).

Now we examine the case	that $\Psi_1=\Psi_2 (\equiv \Psi)$ are infinite	repetition states in which all of their elements $|\psi,i>$	are expressed as in eq.(5).  Then, from	the result in the previous section, it follows that
\begin{equation}
	\rho_{r_N,N}(\Psi) =  |c_N(r_N)|^2
\end{equation}
(As before, when $\Psi_1=\Psi_2=\Psi$, we write	$\rho(\Psi_1,\Psi_2)$ simply as $\rho(\Psi)$.) We can show that its	limit
\begin{equation}
	\rho_r(\Psi) \equiv lim_{N\to\infty}\,N\rho_{r_N,N}(\Psi)
\end{equation}
satisfies the conditions of the	form eqs.(23,24), which	means that, for	any arbitrary positive number of ƒÃ,
\begin{equation}
	\int_{0<r_0-\epsilon} \rho_r(\Psi)\,dr = 0
\end{equation}
\begin{equation}
	\int_{r<r_0+\epsilon} \rho_r(\Psi)\,dr = 1
\end{equation}
(where $r_0 \equiv |a|^2$).  To	prove them, first note that these equations are equivalent to
\begin{equation}
	 lim \sum_{r<r_0-\epsilon} \rho_{r,N}(\Psi) =  0
\end{equation}
\begin{equation}
	 lim \sum_{r>r_0+\epsilon} \rho_{r,N}(\Psi) =  0
\end{equation}
The first identity, for	an example, holds because
\begin{equation}
   \sum_{r<r_0-\epsilon} \rho_{r,N}(\Psi) < 1/\epsilon^2 \sum_{r<r_0-\epsilon} (r_0-r)^2\rho_{r,N}(\Psi) = 1/\epsilon^2 \cdot |a|^2|b|^2/N 
\end{equation}
The right-hand side vanishes in	the limit $N \to \infty$ with ƒÃ fixed.

Now we have understood what we need to derive QPR.  We summarize it as a new postulate of quantum theory as	follows:

\begin{quotation}
Postulate: Suppose that	$\rho_q(\psi_1,\psi_2)$	 be a certain bilinear functional of the quantum state $\psi_1$ and	$\psi_2$ (linear in $\psi_2$ and antilinear in $\psi_1$) which depends on a real variable q (either discrete or	continuous) and reproduces the inner product of the	two states when	they are summed	over all possible values of q
\begin{equation}
    \sum \rho_q(\psi_1,\psi_2)	= <\psi_1|\psi_2>
\end{equation}
Let $\rho_q(\psi)$ be the case in which	$\psi_1=\psi_2=\psi$ with $\psi$ normalized.  If for a certain fixed value $q_0$ and any positive number ƒÃ
\begin{equation}
      \sum_{q<q_0-\epsilon} \rho_q(\psi) =  0
\end{equation}
\begin{equation}
      \sum_{q<q_0+\epsilon} \rho_q(\psi) =  1
\end{equation}
then we	can conclude that the value of q in the	state $\psi$ is	$q_0$. iWhen the variable $q$ is continuous, all the summations above are replaced	with the integrals. Note also that it is not necessary to require that ƒÏ can be defined for arbitrary quantum state ƒÕ.)
\end{quotation}

Note that our $\rho_r(\Psi)$ defined in	eq.(34)	satisfies all the conditions of the above postulate	in the space spanned in	the infinite repetition	states.	 Therefore we conclude that QPR	can be derived from this postulate.

Note also that this postulate supersedes another postulate mentioned in	section 1 about observables.  Information about quantum states should be obtained not from operators but from wave function-based quantities (such as $\rho$).  It does not matter which method	we adopt if both procedures can	be shown to be equivalent, which is the	case in	ordinary applications of quantum mechanics.  But it does not seem to be the	case for the relative frequency	of infinite repetition states.

\section{Measurements of Many but Finite times}

The discussion above tells us what the relative	frequency of a certain individual state should be in an infinite repetition	state.	If accurate measurement	could be carried out for all the infinite individual states, then it would also tell us what the relative frequency	of a certain outcome should be.	 In practice, however, we can not carry	out measurements infinite times.  For the outcome of measurement of, say, a	hundred	times the above	argument itself	does not tell us anything specific.  If	the result is very different from the prediction for the infinite measurement, we would feel uncomfortable and would think that something might be wrong. But can we say something more quantitative about that?

Let us consider an infinite repetition of a hundred measurement.  We already know	that a hundred repetition state	can be expanded	as in eq.(9), which we rewrite here	as
\begin{equation}
	|\Psi>_{100}\,	=  \sum_{n} c_{100}(n)\, |n>_{100}
\end{equation}
where $N=100$ and $n=rN$ is an integer from 0 to 100.  If we apply QPR here, then we would tells that the probability that $|\alpha>$ is observed n	times (out of a	hundred	measurements) is $|c_{100}(n)|^2$.  In our approach in which QPR is replaced with the postulate of the previous section, we	have to	consicer an infinite repetition	of a hundred measurements.  We can tell	that, when a hundred measurements is repeated infinite times, then the relative frequency of the case that $|\alpha>$ is observed n	times (out of 100 measurements)	is $|c_{100}(n)|^2$.  (To prove this statement we should generalize	the two-level system in	section	2 to the 101-level system, which is done in Appendix 1.)	How small (large) it is determines how uncomfortable (or comfortable) we should feel when we get a certain outcome for a hundred measurement.

\section{Philosophical Aspects}

Classical mechanics is regarded	as a theory of realism and determinism,	while quantum mechanics has	been often formulated as a theory of positivism	and probability.  Many people (including Einstein) has felt it unsatisfactory that fundamental laws	of the Nature do not appear to describe	the reality itself, but	rather concerns	about human perception of the Nature. 

There are two aspects in quantum theory	which cause us to suspect that quantum mechanics may not a theory about reality. They are QPR and the reduction of the state.  If the frequencists'	approach can be	justified as has been advocated here,  every reference to the probability in quantum theory	is replaced with the word "frequency".	We can totally get rid of the probability from the theory.  It has also been pointed out in	the context of the many	world interpretation that we can get rid of the	reduction of the state with the	help of	decoherence.

If this	program	turns out to be	successful, its	philosophical implications, especially its ontological meanings should be profound.	 We will come to be able to claim that quantum mechanics is a theory about the  (not hidden but manifestly overt) reality.	 However we should take	note that this reality can not be that of classical physics but	that of	quantum	mechanics.  This should be the reality which can be	superposed and coexist in multiples.  The interpretation of the	reality	as such	might even justify the postulate proposed in this paper, but we would like to refrain from speculating about philosophical issues further here.

\begin{center}
Acknowledgement
\end{center}
The author is indebted to Colin	Bruce who brought his attention	to ref.5).

\begin{flushleft}
Appendix 1: General Case
\end{flushleft}

In section 2 we	presented an example of	the decomposition of the infinite repetition states	in terms or fixed frequency states by using the	two-level system.  Here	we will	show that the calculation can be generalized to	general	multi-level cases.

We are dealing with a symmetric	component of the repetition state and so it is convenient to work with the bosonic creation	operators such as
\[   \alpha^+|0> = |\alpha>,  etc.\]
Then in	our two-level system in	section	2 the N-copy state with	a fixed	relative frequency $r$ of is written as
\[   |r> = (1/\sqrt{n!}\sqrt{(N-n)!})\,\alpha^{+n}\beta^{+N-n} |0> \]
and so the repetition state $|\Psi>_N$ is expanded as
\[ |\Psi>_N=1/\sqrt{N!}\,(a\alpha^++b\beta^+)^N|0>=\sum	1/\sqrt{N!}\,_NC_na^nb^{N-n}\alpha^{+n}\beta^{+N-n}|0> \]
\[=\sum\sqrt{N!/n!(N-n)!}\,a^nb^{N-n}|r>  \]
which reproduces eqs.(9,10).

We generalize to the case in which the single particle state with $M$ levels is
\[  |\psi>\, =\, \sum_{i=1}^M a_i\alpha_i^+ |0>	 \]
This is	normalized as
\[ \sum	|a_i|^2	= 1  \]

Next we	consider the $N$-copy state in which each state	$\alpha_i^+|0>$	have a fixed relative frequency $r_i$ ($\sum r_i=1$	or alternatively $\sum n_i = N$	where $r_i=n_i/N$)
\[  |\{r_i\}> =	\prod_i	1/\sqrt{n_i!}\,	\alpha_i^{+n_i}	|0>  \]
Then the $N$-copy repetition state is decomposed as
\[  |\Psi>_N \equiv 1/\sqrt{N!}\,(\sum_i a_i\alpha_i^+)^N |0>=\sum_{\{n_i\}} 1/\sqrt{N!}\,_NC_{\{n_i\}} \prod_i a_i^{n_i}\alpha_i^{+n_i}|0>	\]
\[ = \sum \sqrt{N!/\prod n_i!}\,\prod a_i^{n_i}|\{r_i\}> \]
where we used the notation
\[    _NC_{\{n_i\}} = N!/\prod_i n_i!  \]

What corresponds to $c_N(r)$ in	the two-level case is
\[ c_N(\{n_i\})	= \sqrt{N!/\prod n_i!}\,\prod a_i^{n_i}	 \]
Then it	is straightforward to prove that
\[  \sum_{\{n_i\}} |c_N(\{n_i\})|^2 = 1	\]
\[  \sum_{\{n_i\}} (r_{i_0}-|a_{i_0}|^2)^2\, |c_N(\{n_i\})|2 = |a_{i_0}|^2\,(1-|a_{i_0}|^2)/N  \]
where $i_0$ is any fixed integer between 1 and $M$ and $r_{i_0}=n_{i_0}/N$.  This means that 
\[  \rho_{r_{i_0},N}(\Psi) =  \sum_{n_{i_0} fixed} \,N\,|c_N(\{n_i\})|2	\]
behaves	like $\delta(r_{i_0}-|a_{i_0}|^2)$ in the limit	$N\to\infty$. Then $\rho_{r_{i_0}}$	is defined as the limit	$N\to\infty$ of	$\rho_{r_{i_0},N}$ and can be used to obtain the required prediction $r_{i_0}=|a_{i_0}|^2$, as was done in section 3 .

\begin{flushleft}
Appendix 2: Continuous Variables
\end{flushleft}

We briefly mention the continuous variable case, taking	quantum	mechanics in one-dimensional space as an example.  First we define a single particle state
\[  |\psi>\, =\, \int \psi(x) A^+(x) |0>	 \]
where $A^+(x)$ is an operator which creates a particle at x. The c-number wave function $\psi(x)$  is	normalized as
\[   \int |\psi(x)|^2 = 1 	 \]

Let $\Delta$ be a region in the x space and define	
\[   \int_{\Delta} |\psi(x)|^2 \equiv |a|^2 	 \]
\[   \int_{\Delta_c} |\psi(x)|^2 \equiv |b|^2 (=1-|a|^2) 	 \]

The N-copy state  
\[  |\Psi>_N =	1/\sqrt{N!}\,\{\int \psi(x) A^+(x)\}^N	|0>  \]
is a superpostion of states each of which has a	certain	number, say n, of particles inside $\Delta$ and other particles outside it (inside $\Delta_c$).  $H(N,n)$ be a space of such states, and	$P_{N,n}$ be a projection operator which projects  $|\Psi>_N$ on	$H(N,n)$.	Then
\[      _N<\Psi|\Psi>_N = \sum_n\,  _N<\Psi|P_{N,n}|\Psi>_N  = 1  \]
Some calculation tells us that 
\[      _N<\Psi|P_{N,n}|\Psi>_N  =\, _NC_n |a|^{2n} |b|^{2(N-n)}   \]
This expression is nothing but the square of eq.(10) and so we can conlude that, in the	limit $N	\to \infty,  r=n/N=|a|^2$, as is expected.  Note that this@method can also be applied to discreet multi-level cases.

\begin{flushleft}
References

1) D.Finkelstein: Transactions of the New York Academy of Sciences \textbf{25}, 621 (1963).

2) J.B.Hartle: Am. J. Phys. \textbf{36}, 704 (1968).

3) E.Fahri, J.Goldstone, and S.Gutmann:	Ann. Phys. \textbf{192}, 368 (1989).

4) J.Gutmann: Phys. Rev. A \textbf{52},	3560 (1995).

5) C.M.Caves and R.Schack: quant-ph/0409144v3 (2005).

6) S.Wada: Butsuri \textbf{44},	(1989) 393 [in Japanese].

\end{flushleft}

\end{document}